\documentclass [aps,showpacs,prd,preprint]{revtex4}
\usepackage[dvips]{graphicx}
\usepackage{bm}
\usepackage{amsmath}

\begin{document}

\title{The Three-body Force and the Tetraquark Interpretation of Light
  Scalar Mesons}

\author{Fan-Yong Zou} 
\affiliation{Department of Physics, Peking University, Beijing 100871,
  China} 

\author{Xiao-Lin Chen} 
\affiliation{Department of Physics, Peking University, Beijing 100871,
  China}

\author{Wei-Zhen Deng} 
\email{dwz@th.phy.pku.edu.cn}
\affiliation{Department of Physics, Peking University, Beijing 100871,
  China}

\begin{abstract}
  We study the possible tetraquark interpretation of light scalar
  meson states $a_0(980)$, $f_0(980)$, $\kappa$, $\sigma$ within the
  framework of the non-relativistic potential model.  The wave
  functions of tetraquark states are obtained in a space spanned by
  multiple Gaussian functions.  We find that the mass spectra of the
  light scalar mesons can be well accommodated in the tetraquark
  picture if we introduce a three-body quark interaction in the quark
  model.  Using the obtained multiple Gaussian wave functions, the decay
  constants of tetraquarks are also calculated within the ``fall apart''
  mechanism.
\end{abstract}


\maketitle

\section{Introduction}
\label{sec:1}

Tetraquarks were proposed decades ago. Early in 1977, Jaffe make a
calculation using the color-magnetic interaction in the bag model
\cite{Jaffe:1976ig, Jaffe:1976ih}. He suggested that the light scalar
mesons below 1 GeV, $a_0(980)$, $f_0(980)$, $\kappa$ and $\sigma$, 
be interpreted as a nonet of light tetraquarks.

In recent years, the light scalar mesons are observed in decays of
charmed mesons.  The $\sigma$ is observed as a peak in decay
$D^+\to\pi^+\pi^-\pi^+$ \cite{Aitala:2000xu,Link:2003gb} and $\kappa$
in $D^+_s\to\pi^+\pi^-\pi^+$ \cite{Aitala:2000xt}.  From the process
$J/\Psi\to\omega\pi^+\pi^-$, BES collaboration determined the pole
position of $\sigma$ to be $M-i\Gamma/2=(541\pm39)-i(252\pm42)$
MeV\cite{Ablikim:2004qn}. Also BES collaboration found a $\kappa$ like
structure in the decay $J/\Psi\to
\bar{K}^*K^+\pi^-$\cite{Ablikim:2005ni}.  The accumulation of
experimental data allows us to study the structure of the light scalar
nonet based on their decay properties
\cite{Maiani:2004uc,Bugg:2006sz,Giacosa:2006tf}.

As a many-body system, a tetraquark state is quite different from a
baryon or a conventional $q\bar{q}$ meson. The color structure is no
longer trivial. It is quite sensitive to the hidden color structure of
QCD interaction.  A tetraquark state, if its existence is confirmed,
may provide us important information about the QCD interaction that is
absent from the ordinary baryons or the $q\bar{q}$ mesons. For
instance, some authors had investigated the tetraquark system with the
three-body $qq\bar{q}$ and $q\bar{q}\bar{q}$ interaction, whose
existence has no direct effect on the ordinary hadron states
\cite{Dmitrasinovic:2003cb,Pepin:2001is,Janc:2004qn}.  The newly
updated experimental data can shed more light on the relation of the
possible tetraquark states and QCD interaction.

In this article, we will study the possible tetraquark state within
the framework of the non-relativistic potential model. We will
calculate mass spectra and wave functions of the light tetraquark
using the Bhaduri potential\cite{Bhaduri:1981pn}.  To fit the
experimental masses, we will extend the model with the three-body
$qq\bar{q}$ and $q\bar{q}\bar{q}$ interaction. Using the wave
functions of tetraquarks, we will determine the coupling constants of
tetraquarks to mesons under the ``fall apart'' mechanism.

The article is organized as follows: In Sec.~\ref{sec:2}, we introduce
the model Hamiltonian and the multiple Gaussian function method which
is used to obtain the tetraquark wave functions.  In Sec.~\ref{sec:3},
we present the ``fall apart'' decay calculation with tetraquark wave
function.  In Sec.~\ref{sec:4}, we present the numerical results.
Finally we will give a brief summary.

\section{Hamiltonian and Wave Functions}
\label{sec:2}

In a non-relativistic quark model, usually the potentials are limited
to the two-body interaction, which mainly consists of two parts: the
$\bm{\lambda}^c\cdot \bm{\lambda}^c$ color interaction including the
confinement and the Coulomb interaction of one-gluon exchange, and
the $\bm{\lambda}^c\cdot \bm{\lambda}^c \bm{\sigma}\cdot\bm{\sigma}$
color-magnetic interaction. The Hamiltonian reads
\begin{equation}
  \label{eq:1}
  H=\sum\limits_{i}(m_{i}+\frac{\bm{P}_i^2}{2m_i})-\frac{3}{4}
  \sum\limits_{i<j}\left[\bm{F}_i\cdot\bm{F}_jV^{C}(r_{ij})+
    \bm{F}_i\cdot\bm{F}_j\bm{S}_i\cdot\bm{S}_jV^{SS}(r_{ij})\right]
\end{equation}
where $m_i$ are quark masses,
$\bm{F}_i^c=\frac{\bm{\lambda}_i^c}2$, and $r_{ij}$ is the distance
between quark $i$ and quark $j$.

Among the various potential forms used in different quark models, the
Bhaduri potential \cite{Bhaduri:1981pn} is rather simple and
gives a unified description of conventional hadron spectroscopy. Also
it is often used to discuss the tetraquark system \cite{Zouzou:1986qh,
  Silvestre-Brac:1993ss, Brink:1998as, Vijande:2003ki, Janc:2004qn}.
The potential reads
\begin{align*}
  V_{ij}^{C}&=-\frac{\kappa}{r_{ij}}+\frac{r_{ij}}{a_{0}^{2}}-D, &
  V_{ij}^{SS}&=\frac{4\kappa}{m_im_j}\frac{1}{r_0^2r_{ij}}e^{-r_{ij}/r_0}.
\end{align*}
The parameter values are
\begin{align}
  \kappa&=102.67\text{Mevfm}, &
  a_0&=0.0326(\text{MeV}^{-1}\text{fm})^{\frac{1}{2}}, 
  &  D&=913.5\text{MeV}, & 
  r_0&=0.4545\text{fm} \notag \\
  m_u&=m_d=337\text{MeV}, & m_s&=600\text{MeV}, & m_c&=1870\text{MeV}, &
  m_b&=5259\text{MeV}.
\end{align}

In a tetraquark, some new interactions which have no direct effect on
the ordinary hadrons may have significant contribution. For instance, one
can introduce the following three-body $qq\bar{q}$ and
$q\bar{q}\bar{q}$
interactions\cite{Dmitrasinovic:2003cb,Pepin:2001is,Janc:2004qn}
\begin{align*}
  V_{qq\bar{q}}(\bm{r}_i,\bm{r}_j,\bm{r}_k) &= d^{abc} F_i^a F_j^b
  F_k^{c*} U_0 \exp[-(r_{ij}^2 + r_{jk}^2 + r_{ki}^2)/r_0^2],\\
  V_{q\bar{q}\bar{q}}(\bm{r}_i,\bm{r}_j,\bm{r}_k) &= d^{abc} F_i^a F_j^{b*}
  F_k^{c*} U_0 \exp[-(r_{ij}^2 + r_{jk}^2 + r_{ki}^2)/r_0^2].
\end{align*}
In this article, since we will only treat the ground states of
tetraquark, the spatial dependence of the three-body interaction is
less important. So we will only add the following simplified
interaction into the model Hamiltonian (\ref{eq:1})
\begin{equation}
  \label{eq:v3b}
  V_{3b} = U_0(d^{abc} F_i^a F_j^b F_k^{c*} +
  d^{abc} F_i^a F_j^{b*} F_k^{c*} ) .
\end{equation}
This interaction is diagonal in the diquark--anti-diquark color base
of tetraquark
\begin{subequations}
  \begin{align}
    \langle [qq]_{\bar{3}} [\bar{q}\bar{q}]_{3} \mid V_{3b} \mid
    [qq]_{\bar{3}} [\bar{q}\bar{q}]_{3} \rangle = - \frac{20}{9}U_0,\\
    \langle [qq]_{6} [\bar{q}\bar{q}]_{\bar{6}} \mid V_{3b} \mid
    [qq]_{6} [\bar{q}\bar{q}]_{\bar{6}} \rangle = + \frac{10}{9}U_0.
  \end{align}
\end{subequations}
An immediate consequence is that this three-body interaction has
no direct contribution to any meson-meson coupling channel.

\begin{figure}[h]
  \begin{equation*}
    \begin{matrix}
      \includegraphics{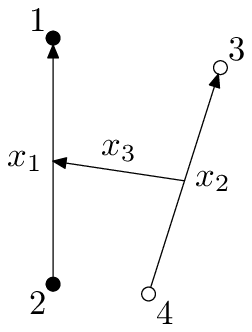} & &
      \includegraphics{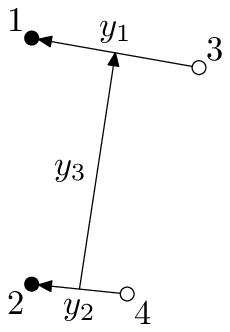} & &
      \includegraphics{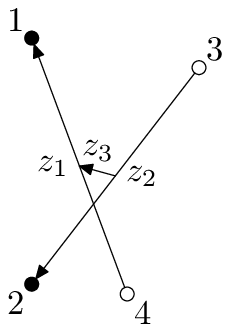} \\
      (a) & \hspace{2cm} & (b) & \hspace{2cm} & (c)
    \end{matrix}
  \end{equation*}
\caption{\label{fig1}Three ways to define the relative coordinates for a
  tetraquark system. Filled and open circles represent quarks and
  anti-quarks respectively.}
\end{figure}

To explain our calculation method, we first define some convenient
coordinates for tetraquark system as illustrated in figure~\ref{fig1}
\cite{Brink:1998as},
\begin{subequations}
  \label{coord-x}
  \begin{align}
    \bm{x}_1 &= \bm{r}_1-\bm{r}_2, \\
    \bm{x}_2 &= \bm{r}_3-\bm{r}_4, \\
    \bm{x}_3 &= \frac{m_1\bm r_1+m_2\bm{r}_2}{m_1+m_2}
    -\frac{m_3\bm{r}_3+m_4\bm{r}_4}{m_3+m_4}
  \end{align}
\end{subequations}
\begin{subequations}
  \label{coord-y}
  \begin{align}
    \bm{y}_1 &= \bm{r}_1-\bm{r}_3, \\
    \bm{y}_2 &= \bm{r}_2-\bm{r}_4, \\
    \bm{y}_3 &= \frac{m_1\bm r_1+m_3\bm{r}_3}{m_1+m_3}
    -\frac{m_2\bm{r}_2+m_4\bm{r}_4}{m_2+m_4}
  \end{align}
\end{subequations}
\begin{subequations}
  \label{coord-z}
  \begin{align}
    \bm{z}_1 &= \bm{r}_1-\bm{r}_4, \\
    \bm{z}_2 &= \bm{r}_2-\bm{r}_3, \\
    \bm{z}_3 &= \frac{m_1\bm r_1+m_4\bm{r}_4}{m_1+m_4}
    -\frac{m_2\bm{r}_2+m_3\bm{r}_3}{m_2+m_3}
  \end{align}
\end{subequations}

The base wave function for the tetraquark will be the product of
color, spin, flavor and spatial wave functions. The color and spin
$SU_c(3) \otimes SU_s(2)$ base function we use is of the following
diquark anti-diquark coupling form:
\begin{itemize}
\item $S=0$
  \begin{align}
    \label{eq:8}
    \alpha_1&=|\bar{3}_{12}3_{34}\rangle_c\otimes|0_{12}0_{34}\rangle_s,
    &
    \alpha_2&=|\bar{3}_{12}3_{34}\rangle_c\otimes|1_{12}1_{34}\rangle_s
    \notag\\
    \alpha_3&=|6_{12}\bar{6}_{34}\rangle_c\otimes|0_{12}0_{34}\rangle_s,
    &
    \alpha_4&=|6_{12}\bar{6}_{34}\rangle_c\otimes|1_{12}1_{34}\rangle_s
  \end{align}
  
\item $S=1$
  \begin{align}
    \alpha_1&=|\overline{3}_{12}3_{34}\rangle_c\otimes|0_{12}1_{34}\rangle_s,
    &
    \alpha_2&=|\overline{3}_{12}3_{34}\rangle_c\otimes|1_{12}0_{34}\rangle_s
    \notag\\
    \alpha_3&=|\overline{3}_{12}3_{34}\rangle_c\otimes|1_{12}1_{34}\rangle_s,
    &
    \alpha_4&=|6_{12}\overline{6}_{34}\rangle_c\otimes|0_{12}1_{34}\rangle_s
    \notag\\
    \alpha_5&=|6_{12}\overline{6}_{34}\rangle_c\otimes|1_{12}0_{34}\rangle_s,
    &
    \alpha_6&=|6_{12}\overline{6}_{34}\rangle_c\otimes|1_{12}1_{34}\rangle_s
  \end{align}

\item $S=2$
  \begin{align}
    \alpha_1&=|\bar{3}_{12}3_{34}\rangle_c\otimes|1_{12}1_{34}\rangle_s,
    &
    \alpha_2=|6_{12}\bar{6}_{34}\rangle_c\otimes|1_{12}1_{34}\rangle_s
  \end{align}
\end{itemize}
Here the color wave function of the two (anti-)quarks is labeled by
$SU_c(3)$ dimension and the spin wave function is by the total spin.

The anti-symmetric diquarks $[ud]$, $[us]$, $[ds]$ form the $\bar{3}$
representation of flavor $SU_f(3)$.  The $\bar{3}$ diquarks and $3$
anti-diquarks further form a tetraquark nonet. They are assumed to be
the light scalar mesons \cite{Jaffe:1976ig, Alford:2000mm,
  Maiani:2004uc}. So the flavor wave functions are:
\begin{subequations}
  \begin{align}
    a_0(I=1,I_3=0)&=\frac1{\sqrt2}([us][\bar{u}\bar{s}]-[ds][\bar{d}\bar{s}])\\
    f_0(I=0) &= \frac{1}{\sqrt{2}}([us][\bar{u}\bar{s}]+[ds][\bar{d}\bar{s}])\\
    \sigma_0(I=0) &= [ud][\bar{u}\bar{d}] \\
    \kappa^+ &= [ud][\bar{s}\bar{d}]
  \end{align}
\end{subequations}

As for the spatial wave functions, we will start from the
multi-dimensional Gaussian function
\begin{equation}
  \label{eq:12}
  g^s(\bm{x}_1,\bm{x}_2,\bm{x}_3)=
  \exp\left(-\sum_{i,j}^3 A_{ij}^s \bm{x}_i\cdot\bm{x}_j \right),
\end{equation}
where $A_{ij}^s$ are the function parameters. The wave function of
this form is well convergent and there exists many analytical
expressions for different matrix elements. We will use it to construct
the spatial wave function base \cite{Suzuki:1998bn,
  SilvestreBrac:2007sg}.

Under the hypothesis of Jaffe, the color-spin wave function of
``good'' diquark is the symmetric one
$|\bar{3}_{12}\rangle_c\otimes|0_{12}\rangle_s$. As the flavor wave
function of the scalar nonet state is anti-symmetric, so the spatial wave
function should be symmetric. That is the spatial wave function of
tetraquark state should be invariant under $\bm{x}_1\to -\bm{x}_1$
and/or $\bm{x}_2\to -\bm{x}_2$. If we use the Gaussian function
(\ref{eq:12}) as the base wave function, it is easy to see that
\cite{Brink:1998as}
\begin{equation*}
  A_{12}=A_{23}=A_{31}=0 .
\end{equation*}

We will use the following symmetric combination as the base
function
\begin{align}
  \label{eq:13}
  \psi^s(\bm{x}_1,\bm{x}_2,\bm{x}_3)&=\frac14[g^s(\bm{x}_1,\bm{x}_2,\bm{x}_3)
  +g^s(-\bm{x}_1,\bm{x}_2,\bm{x}_3)\notag\\
  &\qquad+g^s(\bm{x}_1,-\bm{x}_2,\bm{x}_3)+g^s(\bm{x}_1,\bm{x}_2,-\bm{x}_3)].
\end{align}
If the non-diagonal parameters $A_{ij}(i\ne j)$ are small,  we have
\begin{align}
  \label{eq:14}
  \psi^s(\bm{x}_1,\bm{x}_2,\bm{x}_3) &\approx \exp\left[
    -(A_{11}^s \bm{x}_1^2+A_{22}^s \bm{x}_2^2+A_{33}^s \bm{x}_3^2)\right]
  \notag \\
  &\times \left[1+2A_{12}^{s2}(\bm{x}_1\cdot\bm{x}_2)^2
  +2A_{13}^{s2}(\bm{x}_1\cdot\bm{x}_3)^2
  +2A_{23}^{s2}(\bm{x}_2\cdot\bm{x}_3)^2 \right].
\end{align}
This allow us to study the correlations in the quark alignment.

We will choose $n$ independent symmetric Gaussian functions
(\ref{eq:13}), $s=1,2,...,n$, to span an $n$-dimensional nonorthogonal
basis. The $n$ independent Gaussian functions are obtained by the
following process. First, we use one such symmetric Gaussian function
as the test wave function in variation to determine a base parameter set
$A_{ij}$. The matrix $(A_{ij})$ will be specified by three principal
values denoted $A^{(0)}_{11}$, $A^{(0)}_{22}$, $A^{(0)}_{33}$ and
three Euler angels $(\alpha,\beta,\gamma)$ which specified the
orientation.  Then a complete parameter set $A_{ij}^s(s=1,2,...,n)$
is generated by first scaling to the principal
values\cite{Brink:1998as}
\begin{equation}
  A^{s(0)}_{ii} = A^{(0)}_{ii} d^{s_i}
\end{equation}
where $s_i=-k,-k+1,...,k-1,k$, $(2k+1)^3=n$, and $d$ a scaling
factor; Then we make an Euler rotation $(\alpha,\beta,\gamma)$.

By diagonalizing the Hamiltonian in the above nonorthogonal basis,
we will obtain the mass and wave function of tetraquark states. The
wave function can be expressed in the above base functions as
\begin{equation}
  \label{T-wvf}
  |T\rangle=\phi_f\sum\limits_{i s}C_{i s} \alpha_i \psi^s
\end{equation}
where $\phi_f$ is the flavor wave function, $C_{is}$ is the
superposition coefficient.

Similar to case in pseudo-scalar mesons, the $I=0$ members $f_0$,
$\sigma_0$ in the scalar nonet will mix with each other.  To consider
the mixing, we further introduce a mixing angle $\phi$
\cite{Bugg:2006sz}
\begin{align}
  \label{eq:17}
  f&=f_0\cos\phi+\sigma_0\sin\phi, &
  \sigma&=-\sin\phi f_0+\cos\phi\sigma_0.
\end{align}
Then $f$ and $\sigma$ are the physically observable states. 
In this article, we do not discuss the underlying mechanism of this
mixing. So we will merely treat the mixing angle $\phi$ as one
additional parameter.

\section{Decay Property of Tetraquark State}
\label{sec:3}

Several authors have used the effective Lagrangian with $SU_f(3)$
symmetry to discuss the decay of light scalar nonet
\cite{Maiani:2004uc, Bugg:2006sz}.  Here we can calculate the coupling
constants using the tetraquark wave functions. The general coupling
Lagrangian reads
\begin{align}
  \mathcal{L} &= f_0 \left[
    g_{f_0\pi\pi} \frac{\pi\cdot\pi}{2} 
    + g_{f_0\bar{K}K} \bar{K}{K} + ...
  \right] \notag \\
  &+ \sigma_0\left[
    +g_{\sigma_0\pi\pi} \frac{\pi\cdot\pi}{2} 
    + g_{\sigma_0\bar{K}K} \bar{K}K + ...
  \right] \notag \\
  &+ a \cdot \left[ g_{a_0 \bar{K}{K}} \bar{K}\tau K +
    g_{a \eta_s \pi}
    \eta_s \pi + c_{a\eta_q\pi} \eta_q \pi + ... \right] \notag \\
  &+ g_{\kappa\bar{K}\pi} 
  \left( \bar{K}\tau \kappa \cdot \pi + h.c. \right)
  + ...
\end{align}

\begin{figure}[h]
  \begin{center}
    \includegraphics{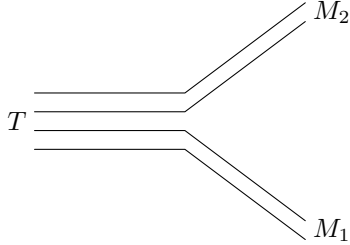}
  \end{center}
  \caption{\label{fig2}``fall apart'' mechanism for decays of
    $q^2\bar{q}^2$ tetraquark states}
\end{figure}

At present, the quark interaction underlying those meson decaying
couplings is still unclear to us. Here we will assume that the
decaying is the fusion process and can be depicted by the ``fall
apart'' mechanism in figure~\ref{fig2}. More specific, we
assume that the coupling constant of a tetraquark $T$ to two mesons
$M_1$ and $M_2$ is proportional to the wave function overlapping
\begin{equation}
  g_{TMM} \propto \langle M_1 M_2 \mid T \rangle.
\end{equation}
The meson wave functions will also be approximated by multiple
Gaussian wave functions determined by a similar variation process
\begin{equation}
  \label{M-wvf}
  |M\rangle_{\bm r}=\phi_f \sum_s C_s \psi^s(\bm{r}),
\end{equation}
where $\phi_f$ is meson flavor wave function, and the spatial base
function is
\begin{equation}
  \psi_s(\bm{r}) = e^{-A^s \bm{r}^2}
\end{equation}
A tetraquark system $q_1q_2\bar{q}_3\bar{q}_4$ can fall apart into two
different flavor combinations $q_1\bar{q}_3+q_2\bar{q}_4$ and
$q_1\bar{q}_4+q_2\bar{q}_3$, and the corresponding final meson-meson
states are different
\begin{align}
  |M_1M_2\rangle_1 &= |M_1\rangle_{\bm{y}_1} |M_2\rangle_{\bm{y_2}}, \\
  |M_1M_2\rangle_2 &= |M_1\rangle_{\bm{z}_1} |M_2\rangle_{\bm{z_2}}. 
\end{align}
The spatial wave functions are in the coordinates $\bm{y}_i$ and
$\bm{z}_i$ defined in Eqs.~(\ref{coord-y}) and (\ref{coord-z})
respectively.

In the decay of the light
scalars to pseudo-scalar mesons, we need to consider the
$\eta$-$\eta'$ mixing
\begin{subequations}
  \begin{align}
    \eta &=\cos\theta\eta_q+\sin\theta\eta_s, \\
    \eta'&=-\sin\theta\eta_q+\cos\theta\eta_s,
  \end{align}
\end{subequations}
where $\eta_q=\frac{1}{\sqrt{2}}(u\bar{u}+d\bar{d})$,
$\eta_s=s\bar{s}$ and $\sin\theta=-0.608$ \cite{Amsler:1992wm}.  We
obtain the following expressions for the coupling constants (a
proportional constant is dropped)
\begin{subequations}
  \label{eq:25}
  \begin{align}
    g_{f_0\to\eta\eta}&=
    \sin\theta\cos\theta A_{f_0\to\eta_q+\eta_s}\\
    g_{f_0\to\eta\eta^\prime}&= \frac1{\sqrt2}
    (\cos\theta^2-\sin\theta^2)A_{f_0\to\eta_q+\eta_s}\\
    g_{f_0\to\eta^\prime\eta^\prime}&=
    -\sin\theta\cos\theta A_{f_0\to\eta_q+\eta_s}\\
    g_{f_0\to KK}&= \frac1{\sqrt2} A_{f_0\to K^++K^-} 
  \end{align}
\end{subequations}
\begin{subequations}
  \label{eq:26}
  \begin{align}
    g_{\sigma_0\to\pi\pi}&=
    \frac{\sqrt3}{2} A_{\sigma_0\to \pi^++\pi^-}\\
    g_{\sigma_0\to\eta\eta}&=
    \frac12\cos\theta^2 A_{\sigma_0\to \eta_q+\eta_q}\\
    g_{\sigma_0\to\eta\eta^\prime}&=
    -\frac1{\sqrt2} \sin\theta\cos\theta A_{\sigma_0\to \eta_q+\eta_q}\\
    g_{\sigma_0\to\eta^\prime\eta\prime}&=
    \frac12\sin\theta^2A_{\sigma_0\to \eta_q+\eta_q} 
  \end{align}
\end{subequations}
\begin{subequations}
  \label{eq:27}
  \begin{align}
    g_{a\to\pi\eta}&= \frac1{\sqrt2} \sin\theta A_{a^+_0\to \pi^++\eta_s}\\
    g_{a\to\pi\eta^\prime}&= \frac1{\sqrt2}
    \cos\theta A_{a^+_0\to \pi^++\eta_s}\\
    g_{a\to KK}&=\frac1{\sqrt2} A_{a^+\to K^++\bar{K}^0}
  \end{align}
\end{subequations}
\begin{subequations}
  \label{eq:28}
  \begin{align}
    g_{\kappa\to\pi K}&=\frac{\sqrt3}{2} A_{\kappa^+\to \pi^++K^0}\\
    g_{\kappa\to\eta K}&=
    \frac{1}{2}\cos\theta A_{\kappa^+\to \eta_q}\\
    g_{\kappa\to\eta^\prime K}&=
    -\frac{1}{2}\sin\theta A_{\kappa^+\to K^++\eta_q}
\end{align}
\end{subequations}
Besides the explicit flavor overlapping factors, $A_{T\to MM}$ is the
overlapping of the color, spin and spatial wave function.

After considering the $\sigma$-$f_0$ mixing effect, The coupling
constants $g_{T\to MM}$ for the decays of $\sigma$ and $f_0$ are
further modified to
\begin{subequations}
  \label{eq:29}
  \begin{align}
    g_{f\to MM}&=\cos\phi g_{f_0\to MM}+\sin\phi
    g_{\sigma_0\to MM},\\
    g_{\sigma\to M M}&=-\sin\phi g_{f_0\to MM}+\cos\phi 
    g_{\sigma_0\to M M}
  \end{align}
\end{subequations}

\section{numerical results}
\label{sec:4}

Bhaduri potential gives a unified description of the spectroscopy of
ordinary mesons and baryons. The Hamiltonian (\ref{eq:1}) itself is an
eigenvalue problem of the differential equation which can be solved
numerically. However, the multiple Gaussian function method can still
give an impressively good approximation of the ground state mesons and
the Gaussian wave function is rather simple to use. In
Table~\ref{tab:1}, we show some results of the pseudo-scalar meson
calculation. We can see that the multiple Gaussian function method
greatly improve the single Gaussian function approximation.
\begin{table}[!h]
  \begin{ruledtabular}
    \begin{tabular}{lccc}
      $m_\pi=m_{\eta_q}$ (MeV) & 136 & 250 & 137\\
      $m_K$ (MeV) & 520 & 582 & 521 \\
      $m_{\eta_s}$ (MeV) & 758 & 800 & 758
    \end{tabular}
  \end{ruledtabular}
  \caption{Pseudo-scalar meson calculation. In col. 1, we direct solve the 
    Schr\"odinger equation. In col. 2, we use the variation method with a 
    single Gaussian function. In col. 3, we use the multiple Gaussian function
    method with 7 Gaussian functions.}
  \label{tab:1}
\end{table}

Now we turn to the tetraquark calculation. In our calculation, the
scaling factor is fixed to be $d=2$. We will take $k=1$, i.e., the
wave function space is spanned by $3^3=27$ Gaussian functions.  In the
light scalar tetraquark, as we assume that the flavors of diquark and
anti-diquark are antisymmetric and the spatial wave function is
symmetric, so the color and spin wave function must be the symmetric
$\alpha_1$ and $\alpha_4$ in eq.~(\ref{eq:8}). 

First, we will consider the original Bhaduri potential without the
three-body quark interaction (\ref{eq:v3b}).  We obtain the following 
masse values
\begin{align}
  M_\sigma &= 687\text{MeV},  & M_\kappa&=1067\text{MeV}  &
  M_{a_0}=M_{f_0} = 1371\text{MeV}.
\end{align}
We can see that the masse values are about 300 MeV higher than the
experimental values.  We can calculate the possibility of a tetraquark
state $|\Psi\rangle$ in different color-spin structure $\alpha_k$
\begin{equation}
  P_{\alpha_k}=\int\prod_{i=1}^{3}d\bm{x}_i \left|\langle\alpha_k|\Psi\rangle
    \right|^2 .
\end{equation}
The color-spin contents of the tetraquark nonet without three quark
interaction are presented in Table~\ref{tab:3}.
\begin{table}[!h]
  \begin{ruledtabular}
    \begin{tabular}{l c c c}
      & $\sigma$ & $\kappa$ & $a_0$, $f_0$ \\
      $P_{\alpha_1}$ & 0.30 & 0.30 & 0.29 \\
      $P_{\alpha_4}$ & 0.70 & 0.70 & 0.71
    \end{tabular}
  \end{ruledtabular}
  \caption{\label{tab:3}%
    Contents of tetraquarks without three-body interaction.}
\end{table}
We see that the color content is mainly the $6\times \bar{6}$ component which
disagrees with Jaffe's ``good'' diquark hypothesis.

Next, we will turn on the three body interaction with $U_0=0.333$GeV.
We find that light tetraquark masses are
\begin{align}
  M_\sigma &= 443\text{MeV},  & M_\kappa&=744\text{MeV}  &
  M_{a_0}=M_{f_0} = 985\text{MeV},
\end{align}
which are in agreement with the experimental values\cite{Yao:2006px}:
\begin{align}
  M_\sigma &= 800\pm400\text{MeV},  & M_\kappa&=840\pm80\text{MeV}, 
  \notag\\
  M_{a_0}&=984.7\pm1.2\text{MeV}, & M_{f_0} &= 980\pm10\text{MeV}.
\end{align}
The color-spin contents of the nonet are shown in in Table~\ref{tab:5}
which also agree with the ``good'' diquark
picture\cite{Maiani:2004uc,Maiani:2004vq}.
\begin{table}[!h]
  \begin{ruledtabular}
    \begin{tabular}{l c c c}
      & $\sigma$ & $\kappa$ & $a_0$, $f_0$ \\
      $P_{\alpha_1}$ & 0.80 & 0.88 & 0.92 \\
      $P_{\alpha_4}$ & 0.20 & 0.12 & 0.08
    \end{tabular}
  \end{ruledtabular}
  \caption{\label{tab:5}%
    Contents of tetraquarks with three-body interaction.}
\end{table}

In our calculation, the tetraquark wave function is symmetric under
the coordinates reflections $\bm{x}_1\to-\bm{x}_1$ and/or
$\bm{x}_2\to-\bm{x}_2$. It is easy to see that the expectation
values
\begin{equation}
  \langle \bm{x}_i\cdot\bm{x}_j \rangle = \langle \bm{x}_i^2 \rangle
  \delta_{ij}.
\end{equation}
$\sqrt{\langle \bm{x}_1^2 \rangle}$ and $\sqrt{\langle \bm{x}_2^2
  \rangle}$ are the radii of mean square (RMS) of diquark and
anti-diquark respectively. The quark and anti-quark RMS in tetraquark
is
\begin{align}
  R^2 &\equiv \frac{\langle \sum_{i=0}^4 
    m_i(\bm{r}_i-\bm{R}_{\text{CM}})^2\rangle}{\sum_{i=0}^4 m_i}
  \notag \\
  &= \frac{\mu_{12} \langle \bm{x}_1^2 \rangle
  + \mu_{34} \langle \bm{x}_2^2 \rangle
  + \mu_{12,34}
  \langle \bm{x}_3^2 \rangle}{m_1+m_2+m_3+m_4},
\end{align}
where
\begin{align}
  \bm{R}_{\text{CM}} &= \frac{\sum_{i=0}^4 m_i \bm{r}_i}{\sum_{i=0}^4 m_i},\\
  \mu_{ij} &= \frac{m_i m_j}{m_i + m_j}, \\
  \mu_{ij,kl} &= \frac{(m_i+m_j)(m_k+m_l)}{m_i+m_j+m_k+m_l}.
\end{align}
The RMS values are tabulated in Table~(\ref{tab:6}).
\begin{table}[!h]
  \begin{ruledtabular}
    \begin{tabular}{c c c c}
      & $\sigma$ & $\kappa$ & $a_0$, $f_0$ \\
      $\sqrt{\langle \bm{x}_1^2 \rangle}$ & 0.70 & 0.72 & 0.70 \\
      $\sqrt{\langle \bm{x}_2^2 \rangle}$ & 0.70 & 0.69 & 0.70 \\
      $\sqrt{\langle \bm{x}_3^2 \rangle}$ & 0.54 & 0.58 & 0.56 
    \end{tabular}
  \end{ruledtabular}
  \caption{\label{tab:6}%
    The RMS values in fm.}
\end{table}

However, the spatial wave function (\ref{eq:13}) is
beyond the usual tetraquark assumption. Usually a tetraquark
is assumed to be constructed from the ``good diquark''. The inner
orbital angular momentum of the (anti-)diquark in a tetraquark is
zero.  So the relative angular momentum between the diquark and
anti-diquark in the scalar tetraquark is also zero. The spatial wave
function will has the form
\begin{equation}
  \label{eq:38}
  \psi(\bm{x}_1,\bm{x}_2,\bm{x}_3)=\psi(\bm{x}_1^2,\bm{x}_2^2,\bm{x}_3^2),
\end{equation}
i.e., all the $\bm{x}_i$ are in S-waves.  Our choice (\ref{eq:13}) is
beyond the above assumption (this can be easily see from
eq.~(\ref{eq:14})).  If eq.~(\ref{eq:38}) holds, we will have
following identity
\begin{equation}
  \langle (\bm{x}_i\cdot \bm{x}_j)^2 \rangle = \frac13
  \langle \bm{x}_i^2 \rangle\langle \bm{x}_j^2 \rangle
  \qquad (i\ne j).
\end{equation}
We will measure the deviation from (\ref{eq:38}) of a tetraquark state
by
\begin{equation}
  \epsilon_{ij} = \frac{3\langle (\bm{x}_i\cdot \bm{x}_j)^2 \rangle} 
  {\langle \bm{x}_i^2 \rangle\langle \bm{x}_j^2 \rangle} - 1.
\end{equation}
The numerical $\epsilon_{ij}$ values are listed in Table~\ref{tab:7}.
The small nonzero $\epsilon$ values means that the tetraquark states
are indeed not pure in S-wave.  There is always some D-wave mixing.
\begin{table}[!h]
  \begin{ruledtabular}
    \begin{tabular}{c c c c}
      & $\sigma$ & $\kappa$ & $a_0$, $f_0$ \\
      $\epsilon_{12}$ & 0.14 & 0.23 & 0.08 \\
      $\epsilon_{13}$ & 0.21 & 0.22 & 0.13 \\
      $\epsilon_{23}$ & 0.21 & 0.24 & 0.13 
    \end{tabular}
  \end{ruledtabular}
  \caption{\label{tab:7}%
    The $\epsilon_{ij}$ values of tetraquark wave function.}
\end{table}

With the obtained wave functions, we can calculate the wave function
overlapping in Eqs.~(\ref{eq:25})--(\ref{eq:28}) to get the coupling
constants.  The results are collected in Table~\ref{tab:8}.
\begin{table}[!h]
  \begin{ruledtabular}
    \begin{tabular}{c c c c c}
      ${\sigma_0\to\pi^++\pi^-}$ & ${\kappa^+\to\pi^++K^0}$  
      & ${\kappa^+\to K^++d\bar{d}}$  & ${a^+\to K^++\bar{K}^0}$  
      & ${a^+\to\pi^++\eta_s}$\\
      10.75   &9.37     &9.37    &8.16   &8.38
    \end{tabular}
  \end{ruledtabular}
  \caption{\label{tab:8}%
    Tetraquark--meson-meson wave function overlapping in color, spin, 
    spatial space $A_{T\to M_1M_2}$ (unit $\text{GeV}^{-3/2}$).}
\end{table}
According to ref.~\onlinecite{Bugg:2006sz}, the scalar isoscalar
mixing angle $\phi$ in eq.~(\ref{eq:17}) will be fixed by the ratio
$g^2_{f\to\bar{K}K}/g^2_{f\to\pi\pi}=4.21$ with eq.~(\ref{eq:29}).
This gives $\phi=16.8^\circ$.  The ratios of coupling constants for
scalar meson decays are listed in Table~\ref{tab:9}. Similar to Bugg's
calculation\cite{Bugg:2006sz}, although most of the experimental
ratios can be fitted within a factor $2$,
$g^2_{f\to\eta\eta}/g^2_{f\to\pi\pi}$ is far above experimental value.
\begin{table}[!h]
  \begin{ruledtabular}
    \begin{tabular}{c c c c c}
      &Analysis of ref.~\onlinecite{Giacosa:2006rg} 
      &Analysis of ref.~\onlinecite{Bugg:2006sz}  
      &our results  
      &Expt\cite{Bugg:2005nt,Bugg:2006sr,Bugg:2005xx,Ablikim:2004qn}\\
      $g^2_{a_0\to\pi\eta}/g^2_{a_0\to\bar{K}K}$          
      &0.60  &$0.40\pm0.03$ &0.39  &$0.75\pm0.11$\\
      $g^2_{f\to\bar{K}K}/g^2_{f\to\pi\pi}$
      &4.21  &$4.21\pm0.46$ &4.21  &$4.21\pm0.46$\\
      $g^2_{f\to\bar{K}K}/g^2_{a_0\to\bar{K}K}$ 
      &2.28  &$0.93\pm0.01$ &0.92  &$2.15\pm0.4$\\
      $g^2_{a_0\to\pi\eta^\prime}/g^2_{a_0\to\pi\eta}$
      &0.16  &-  &1.71  &-\\
      $g^2_{f\to\eta\eta}/g^2_{f\to\pi\pi}$
      &1.35  &$1.07\pm0.18$  &1.15  &$<0.33$\\
      $g^2_{\sigma\to\bar{K}K}/g^2_{\sigma\to\pi\pi}$ 
      &$4.8\times 10^{-7}$  &$0.03\pm0.01$  &0.04  &$0.6\pm0.1$\\
      $g^2_{\sigma\to\eta\eta}/g^2_{\sigma\to\pi\pi}$     
      &0.05  &$0.23\pm0.02$  &0.25  &$0.20\pm0.04$\\
      $g^2_{\kappa\to\pi K}/g^2_{\sigma\to\pi\pi}$
      &0.78  &0.58 &0.83  &$(2.14\pm0.28)$ to $(1.35\pm0.10)$\\
      $g^2_{\kappa\to\eta K}/g^2_{\kappa\to\pi K}$
      &0.12  &$0.20\pm0.01$ &0.21  &$0.06\pm0.02$\\
      $g^2_{\kappa\to\eta^\prime K}/g^2_{\kappa\to\pi K}$ 
      &0.006  &$0.13\pm0.01$ &0.12  &$0.29\pm0.29$
    \end{tabular}
  \end{ruledtabular}
  \caption{\label{tab:9}%
    Ratios of coupling constants for light scalar meson decays, 
    with $\phi=16.8^\circ$.}
\end{table}

\section{Summary}
\label{sec:5}

In summary, we have performed a tetraquark calculation of light scalar
mesons using the quark potential model. If we only consider the
two-body quark interaction as in the conventional hadron calculation,
the masses of the tetraquark states will be several hundred MeV higher
than experimental data. Also the major component of the light
tetraquark wave functions consists of the color sextet diquark and
anti-diquark.  After including a three-body interaction in the
Hamiltonian, we find the masses of the light tetraquark nonet agree
with experimental data and their wave functions are composed of mainly
the ``good'' diquark and anti-diquark. We have used the multiply
Gaussian function to approximate the tetraquark wave functions and
noticed that there is a small mixing of D-waves in the wave functions.
With wave functions we obtain, we also calculate the coupling
constants for scalar meson decays according to the ``fall apart''
mechanism. By introducing the isoscalar mixing angle $\phi$, we obtain
a fit of the ratios of coupling constants for scalar meson decays
similar to other analysis based on tetraquark picture.

\begin{acknowledgments}
  We would like to thank Zhan Shu, Yan-Rui Liu, and Shi-Lin Zhu for
  useful discussions. This work was supported by the National Natural
  Science Foundation of China under Grants 10675008.
\end{acknowledgments}


\begin{thebibliography}{10}
\providecommand{\url}[1]{\texttt{#1}}
\providecommand{\urlprefix}{URL }
\expandafter\ifx\csname urlstyle\endcsname\relax
  \providecommand{\doi}[1]{doi:\discretionary{}{}{}#1}\else
  \providecommand{\doi}{doi:\discretionary{}{}{}\begingroup
  \urlstyle{rm}\Url}\fi
\providecommand{\eprint}[2][]{\url{#2}}

\bibitem{Jaffe:1976ig}
R.~L. Jaffe,
\newblock Phys. Rev., 1977,
\newblock \textbf{D15}: 267

\bibitem{Jaffe:1976ih}
R.~L. Jaffe,
\newblock Phys. Rev., 1977,
\newblock \textbf{D15}: 281

\bibitem{Aitala:2000xu}
E.~M. Aitala, et~al. (E791),
\newblock Phys. Rev. Lett., 2001,
\newblock \textbf{86}: 770--774,
\newblock \eprint{hep-ex/0007028}

\bibitem{Link:2003gb}
J.~M. Link, et~al. (FOCUS),
\newblock Phys. Lett., 2004,
\newblock \textbf{B585}: 200--212,
\newblock \eprint{hep-ex/0312040}

\bibitem{Aitala:2000xt}
E.~M. Aitala, et~al. (E791),
\newblock Phys. Rev. Lett., 2001,
\newblock \textbf{86}: 765--769,
\newblock \eprint{hep-ex/0007027}

\bibitem{Ablikim:2004qn}
M.~Ablikim, et~al. (BES),
\newblock Phys. Lett., 2004,
\newblock \textbf{B598}: 149--158,
\newblock \eprint{hep-ex/0406038}

\bibitem{Ablikim:2005ni}
M.~Ablikim, et~al. (BES),
\newblock Phys. Lett., 2006,
\newblock \textbf{B633}: 681--690,
\newblock \eprint{hep-ex/0506055}

\bibitem{Maiani:2004uc}
L.~Maiani, F.~Piccinini, A.~D. Polosa, V.~Riquer,
\newblock Phys. Rev. Lett., 2004,
\newblock \textbf{93}: 212002,
\newblock \eprint{hep-ph/0407017}

\bibitem{Bugg:2006sz}
D.~V. Bugg,
\newblock Eur. Phys. J., 2006,
\newblock \textbf{C47}: 57--64,
\newblock \eprint{hep-ph/0603089}

\bibitem{Giacosa:2006tf}
F.~Giacosa,
\newblock Phys. Rev., 2007,
\newblock \textbf{D75}: 054007,
\newblock \eprint{hep-ph/0611388}

\bibitem{Dmitrasinovic:2003cb}
V.~Dmitrasinovic,
\newblock Phys. Rev., 2003,
\newblock \textbf{D67}: 114007

\bibitem{Pepin:2001is}
S.~Pepin, F.~Stancu,
\newblock Phys. Rev., 2002,
\newblock \textbf{D65}: 054032,
\newblock \eprint{hep-ph/0105232}

\bibitem{Janc:2004qn}
D.~Janc, M.~Rosina,
\newblock Few Body Syst., 2004,
\newblock \textbf{35}: 175--196,
\newblock \eprint{hep-ph/0405208}

\bibitem{Bhaduri:1981pn}
R.~K. Bhaduri, L.~E. Cohler, Y.~Nogami,
\newblock Nuovo Cim., 1981,
\newblock \textbf{A65}: 376--390

\bibitem{Zouzou:1986qh}
S.~Zouzou, B.~Silvestre-Brac, C.~Gignoux, J.~M. Richard,
\newblock Z. Phys., 1986,
\newblock \textbf{C30}: 457

\bibitem{Silvestre-Brac:1993ss}
B.~Silvestre-Brac, C.~Semay,
\newblock Z. Phys., 1993,
\newblock \textbf{C57}: 273--282

\bibitem{Brink:1998as}
D.~M. Brink, F.~Stancu,
\newblock Phys. Rev., 1998,
\newblock \textbf{D57}: 6778--6787

\bibitem{Vijande:2003ki}
J.~Vijande, F.~Fernandez, A.~Valcarce, B.~Silvestre-Brac,
\newblock Eur. Phys. J., 2004,
\newblock \textbf{A19}: 383,
\newblock \eprint{hep-ph/0310007}

\bibitem{Alford:2000mm}
M.~G. Alford, R.~L. Jaffe,
\newblock Nucl. Phys., 2000,
\newblock \textbf{B578}: 367--382,
\newblock \eprint{hep-lat/0001023}

\bibitem{Suzuki:1998bn}
Y.~Suzuki, K.~Varga,
\newblock Lect. Notes Phys., 1998,
\newblock \textbf{M54}: 1--310

\bibitem{SilvestreBrac:2007sg}
B.~Silvestre-Brac, V.~Mathieu, 2007,
\newblock \eprint{arXiv:0706.2300 [hep-ph]}

\bibitem{Amsler:1992wm}
C.~Amsler, et~al. (Crystal Barrel),
\newblock Phys. Lett., 1992,
\newblock \textbf{B294}: 451--456

\bibitem{Yao:2006px}
W.~M. Yao, et~al. (Particle Data Group),
\newblock J. Phys., 2006,
\newblock \textbf{G33}: 1--1232

\bibitem{Maiani:2004vq}
L.~Maiani, F.~Piccinini, A.~D. Polosa, V.~Riquer,
\newblock Phys. Rev., 2005,
\newblock \textbf{D71}: 014028,
\newblock \eprint{hep-ph/0412098}

\bibitem{Giacosa:2006rg}
F.~Giacosa,
\newblock Phys. Rev., 2006,
\newblock \textbf{D74}: 014028,
\newblock \eprint{hep-ph/0605191}

\bibitem{Bugg:2005nt}
D.~V. Bugg, 2005,
\newblock \eprint{hep-ex/0510014}

\bibitem{Bugg:2006sr}
D.~V. Bugg,
\newblock Eur. Phys. J., 2006,
\newblock \textbf{C47}: 45--55,
\newblock \eprint{hep-ex/0603023}

\bibitem{Bugg:2005xx}
D.~V. Bugg,
\newblock Phys. Lett., 2006,
\newblock \textbf{B632}: 471--474,
\newblock \eprint{hep-ex/0510019}

\end{thebibliography}

\end{document}